\documentstyle[12pt, a4]{article}
\begin{document}
\author{B. G\"{o}n\"{u}l
\and Department of Engineering Physics,Faculty of Engineering,\and
University of Gaziantep, 27310 Gaziantep-T\"{u}rkiye}
\title{Exact treatment of $\ell \neq 0$ states}
\date{}
\maketitle
\begin{abstract}
Using the basic ingredient of supersymmetry, a general procedure
for the treatment of quantum states having nonzero angular momenta
is presented.
\end{abstract}

%\section{Introduction}
Over the years the Schr\"{o}dinger equation has been studied extensively
regarding its exact solvability. Many advances have been made
in this area by classifying quantum mechanical potentials according
to their symmetry properties. For instance, various algebra which
reveal the underlying symmetry as well as facilitating obtaining
the solutions have been found. In this respect, the application
of supersymmetry ideas to nonrelativistic quantum mechanics has
revived fresh interest in the problem of obtaining algebraic
solutions of exactly solvable nonrelativistic potentials and
provided a deeper understanding of analytically solvable Hamiltonians,
as well as a set of powerful approximate schemes for dealing
with problems admitting no exact solutions. The concept of shape
invariance [1] has played a fundamental role in these developments.

In this letter, a novel method within the frame of supersymmetric
quantum mechanics [1] is introduced, using the spirit of perturbation
theory, for the treatment $\ell > 0$ states assuming
that the potential of interest is exactly solvable for $\ell =0$ state.

For the consideration of spherically symmetric potentials, the
corresponding Schr\"{o}dinger equation for the radial wave function
reads
\begin{equation}
\frac{\hbar ^{2} }{2m} \frac{\psi''_{n}(r)}{\psi_{n}(r)}=
\left[ V(r)-E_{n} \right] \quad ,\quad V(r)=\left[ V_{0} (r)+\frac{\hbar
^{2} }{2m} \frac{\ell (\ell +1)}{r^{2} } \right]~.
\end{equation}

Now, assuming that the term of angular momentum barrier can be
treated like a perturbing potential, which will be discussed
below, one writes the wave function $\psi _{n} $ as
\begin{equation}
\psi _{n} (r)=\chi _{n} (r)\phi _{n} (r),
\end{equation}
in which $\chi _{n} $ is the known normalized
eigenfunction of the unperturbed Schr\"{o}dinger equation corresponding to
$\ell =0$ state whereas $\phi _{n} $ is a moderating function due to the
angular momentum barrier. Substituting (2) into (1) yields
\begin{equation}
\frac{\hbar ^{2} }{2m} \left( \frac{\chi''_{n}}{\chi _{n} }
+\frac{\phi''_{n} }{\phi _{n} } +2\frac{\chi ' _{n} }{\chi _{n} }
\frac{\phi ' _{n} }{\phi _{n} } \right) =V-E_{n}~.
\end{equation}

With the new definitions,
\begin{equation}
W_{n} =-\frac{\hbar }{\sqrt{2m} } \frac{\chi ' _{n} }{\chi _{n} } \quad
,
\Delta W_{n} =-\frac{\hbar }{\sqrt{2m} } \frac{\phi ' _{n} }{\phi _{n} }~,
\end{equation}
one arrives at
\begin{equation}
\frac{\hbar ^{2} }{2m} \frac{\chi''_{n} }{\chi _{n} } =W_{n} ^{2}
-\frac{\hbar }{\sqrt{2m} } W' _{n} =V_{0} (r)-\epsilon _{n}~,
\end{equation}
where $\epsilon _{n} $ is the eigenvalue for the exactly solvable potential for
$\ell =0$ case, and
\begin{equation}
\frac{\hbar ^{2} }{2m} \left( \frac{\phi''_{n} }{\phi _{n} }
+2\frac{\chi ' _{n} }{\chi _{n} } \frac{\phi ' _{n} }{\phi _{n} } \right)
=\Delta W_{n} ^{2} -\frac{\hbar }{\sqrt{2m} } \Delta W' _{n} +2W_{n}
\Delta W_{n} =\frac{\ell \left( \ell +1\right) \hbar ^{2} }{2mr^{2} }
-\Delta \epsilon _{n}~,
\end{equation}
in which $\Delta \epsilon _{n} $ is the correction term
to the energy due to the barrier term, and $E_{n} =\epsilon _{n} +\Delta \epsilon _{n} $.
Subsequently, Eq. (3) reduces to
\begin{equation}
\left( W_{n} +\Delta W_{n} \right) ^{2} -\frac{\hbar }{\sqrt{2m} } \left(
W_{n} +\Delta W_{n} \right)'=V-E_{n}~,
\end{equation}
from which it is clear that the whole superpotential, $W_{n} +\Delta W_{n}$,
should address exactly solvable potentials. In other words, Eq. (7) is a novel
sophisticated and extended treatment of Eq. (5). That is why the present model
works well for all analytically solvable potentials. For a recent application of
the technique used here regarding perturbed Coulomb interactions the reader is
referred to \cite{ozer}.

In principle as one knows explicitly the solution of (5), namely
the whole spectrum and corresponding eigenfunctions of the potential
$V_{0} (r)$, the goal here is to solve only Eq. (6), which is the main result
of this letter, leading to the full corrections to the energy
and wave functions for all quantum states with $\ell \neq 0$.

To test the effectiveness of the model we consider through this
short letter three well known problems of quantum mechanics as
illustrative examples. We first apply the model to the three
dimensional harmonic oscillator problem
\begin{equation}
V(r)=\frac{mw^{2} r^{2} }{2} +\frac{\ell \left( \ell +1\right) \hbar ^{2}
}{2mr^{2} }~.
\end{equation}

Starting with its exact solutions [1] for $n=\ell =0$,
\begin{equation}
W_{0} (r)=\sqrt{\frac{m}{2} } wr-\frac{\hbar }{\sqrt{2m} r}~,~
\chi_{0} (r)=Nr\exp (-\frac{mwr^{2} }{2\hbar } )~,~
\epsilon_{0} =\frac{3}{2} \hbar w~,
\end{equation}
and introducing a superpotential for the treatment of quantum states,
via Eq. (6), with nonzero angular momenta
\begin{equation}
\Delta W_{0} (r)=-\frac{\ell \hbar }{\sqrt{2m} r}~,
\end{equation}
one easily obtains the full corrections to the wave function
and energy in (9) due to the effect of angular momentum barrier,
\begin{equation}
\phi _{0} (r)=\exp \left( -\frac{\sqrt{2m} }{\hbar }
\int\limits_{}^{r}\Delta W_{0} (z)dz \right) =r^{\ell } \quad ,\quad
\Delta \epsilon _{0} =\ell \hbar w~.
\end{equation}

From which, the exact solution of the potential in (8) reads
\begin{equation}
\psi _{0} (r)=\chi _{0} (r)\phi _{0} (r)=Nr^{\ell +1} \exp \left(
-\frac{mwr^{2} }{2\hbar } \right) \quad ,\quad E_{0} =\left( \ell
+\frac{3}{2} \right) \hbar w~.
\end{equation}

These are indeed the exact results [1] which can be checked out
also by the use of Eq. (7).

We proceed with another example,
\begin{equation}
V(r)=-\frac{e^{2} }{r} +\frac{\ell \left( \ell +1\right) \hbar ^{2}
}{2mr^{2} }~,
\end{equation}
having in mind, as in the previous example, that analytical solutions
of (13) for $n=\ell =0$ is known [1]
\begin{equation}
W_{0} (r)=\sqrt{\frac{m}{2} } \frac{e^{2} }{\hbar } -\frac{\hbar
}{\sqrt{2m} r}~,~\chi _{0} (r)=Nr\exp \left( -\frac{me^{2}
r}{\hbar ^{2} } \right)~,~\epsilon _{0} =-\frac{me^{4} }{2\hbar^{2} }~.
\end{equation}

Setting the superpotential $\Delta W_{0} $
\begin{equation}
\Delta W_{0} (r)=-\frac{\ell \hbar }{\sqrt{2m} r} -\sqrt{\frac{m}{2} }
\frac{\ell e^{2} }{\left( \ell +1\right) \hbar }~,
\end{equation}
and substituting it, together with $W_{0} $ in (14), into (6) yields
\begin{equation}
\Delta \epsilon _{0} =-\frac{me^{4} }{2\hbar ^{2} } \left[
\frac{1}{\left( \ell +1\right) ^{2} } -1\right]~,
\end{equation}
from which, the correct energy is
\begin{equation}
E_{0} =\epsilon _{0} +\Delta \epsilon _{0} =-\frac{me^{4} }{2\hbar ^{2} }
\left[ \frac{1}{\left( \ell +1\right) ^{2} } \right]~,
\end{equation}
and from (15), the moderating function
\begin{equation}
\phi _{0} (r)=\exp \left[ -\frac{\sqrt{2m} }{\hbar } \int\limits_{}^{r}\Delta W_{0}
(z)dz \right] =r^{\ell } \exp \left[ \frac{m\ell e^{2} r}{(\ell +1)\hbar
^{2} } \right]~,
\end{equation}
which leads to the full wavefunction
\begin{equation}
\psi _{0} (r)=\chi _{0} (r)\phi _{0} (r)=Nr^{\ell +1} \exp \left[
-\frac{me^{2} r}{(\ell +1)\hbar ^{2} } \right]~.
\end{equation}

Eqs. (17) and (19) justify once more the reliability of the present
model since they are exact [1].

Finally, we consider the corrected form of a super family
potential [3] which is known in the literature as the approximate
Hulthen effective potential introduced by Greene and Aldrich [4]
in their method to generate pseudo-Hulthen wave function for
states,
\begin{equation}
V_{\ell +1} (r)=\frac{\hbar ^{2} }{2m} \frac{\ell \left( \ell +1\right)
\alpha ^{2} \exp \left( -2\alpha r\right) }{\left[ 1-\exp \left( -\alpha
r\right) \right] ^{2} } -\frac{e^{2} \alpha \exp \left( -\alpha r\right)
}{\left[ 1-\exp \left( -\alpha r\right) \right] } \left[ 1-\ell \left(
\ell +1\right) \frac{\beta }{2} \right]~,
\end{equation}
that (considering $\beta =\alpha \hbar ^{2} /me^{2}$) reduces to
\begin{equation}
V_{\ell +1} (r)=\frac{\hbar ^{2} }{2m} \frac{\ell \left( \ell
+1\right) \alpha ^{2} \exp \left( -\alpha r\right) }{\left[ 1-\exp
\left( -\alpha r\right) \right] ^{2} } -\frac{e^{2} \alpha \exp
\left( -\alpha r\right) }{\left[ 1-\exp \left( -\alpha r\right)
\right] }~.
\end{equation}
In (20), $(\ell +1)$ denotes the partner number with $\ell
=0,1,2,\ldots $ leading to the supersymmetric partner potentials.
It is noted that for $\ell =0$ the potentials in (20) and (21)
lead to the well known Hulthen potential.

Though we know the potential in (20) is exactly solvable [3],
to illustrate again the elegancy of the present treatment, we
assume for a while that the Schr\"{o}dinger equation with this
potential has an analytic solution for only $\ell =0$, and try
to calculate corrections to the solution within the
framework Eq. (6) due to the other states with $\ell > 0$ in the same system.

The superpotential, corresponding ground state wavefunction and
energy expressions for the Hulthen potential are
\begin{eqnarray}
W_{0} (r) & = & \sqrt{\frac{m}{2} } \frac{e^{2} }{\hbar } \left( 1-\beta /2\right)-
\frac{\hbar }{2m} \frac{\alpha \exp \left( -\alpha r\right)}
{\left[ 1-\exp \left( -\alpha r\right) \right] }~,~
\nonumber
\\
\chi _{0}(r) & = & N\left[ 1-\exp \left( -\alpha r\right) \right]
\exp \left[-\frac{me^{2} r}{\hbar ^{2} } \left( 1-\beta /2 \right) \right]~,~
\nonumber
\\
\epsilon _{0} & = & -\frac{me^{4} }{2\hbar ^{2} } \left( 1-\beta /2 \right)^{2}~.
\end{eqnarray}

For the present example, the unique superpotential leading to
the exact corrections should read
\begin{equation}
\Delta W_{0} (r)=-\sqrt{\frac{m}{2} } \frac{\ell e^{2} }{\hbar } \left(
\frac{1}{\ell +1} +\frac{\beta }{2} \right) -\frac{\ell \hbar \alpha \exp
\left( -\alpha r\right) }{\sqrt{2m} \left[ 1-\exp \left( -\alpha r\right)
\right] }~,
\end{equation}
which can be tested through Eq. (6), from which
\begin{equation}
\Delta \epsilon _{0} =-\frac{me^{4} }{2\hbar ^{2} } \left[
\frac{1}{\left( \ell +1\right) } -\frac{\left( \ell +1\right) \beta }{2}
\right] ^{2} +\frac{me^{4} }{2\hbar ^{2} } \left( 1-\beta /2 \right) ^{2}~.
\end{equation}

Thus, one readily sees that
\begin{equation}
\psi _{0} (r)=\chi _{0} (r)\phi _{0} (r)=
N\left[ 1-\exp \left( -\alpha r\right) \right]^{\ell +1}
\exp \left\{ -\frac{me^{2} }{\hbar ^{2} }
\left[ \frac{1}{\ell +1} -\frac{\left( \ell +1\right) \beta }{2} \right] r\right\}~,
\end{equation}
and
\begin{equation}
E_{0} =\epsilon _{0} +\Delta \epsilon _{0} =-\frac{me^{4} }{2\hbar ^{2} }
\left[ \frac{1}{\left( \ell +1\right) } -\frac{\left( \ell +1\right) \beta
}{2} \right] ^{2}~,
\end{equation}
which are exact [3] for the potential in (20).

Although we have considered here only the ground solutions for
the sake of clarity, the application of our simple approach to
excited states does not cause any problem. As the entire spectrum
wave functions for the unperturbed part of the potential of we
interest are known explicitly, what one needs is just to set
the corresponding superpotential via
$W_{n} =-\hbar /\sqrt{2m} \left( \chi ' _{n} /\chi _{n}  \right)$ to use in (6),
together with a properly chosen superpotential leading to the barrier term with
$\ell \neq 0$. In this respect, Eq. (4) in our work put forward a new perspective
when compared to the usual treatment of supersymmetric quantum
mechanics in which generally the superpotentail is related to
the ground state wavefunction. This fresh idea in (4) would also
be helpful in treating excited states within the frame of supersymmetric
perturbation theory [5] where one needs to deal with nasty integrals
and tedious calculation procedures. Furthermore, Eq. (4)
allows us having a unique but closed analytical
expression for the energy corrections involving all states. Examples of such treatmens,
together with a new look at perturbation problems, are discussed
in detail by the present author, which will appear elsewhere.

We believe that the present technique would find a wide application
in the related area. In particular, such treatments would shed
a light in the search of analytical solutions of Morse, Rosen-Morse,
Eckart, P\"{o}schl-Teller and Scarf potentials in case $\ell \neq 0$.
Along this line the works are in progress.

Finally, from the whole discussion presented in this letter it is
obvious that $\Delta \epsilon \rightarrow 0$ and $\Delta
W\rightarrow 0$ in case $\ell \rightarrow 0$, which means that
$\ell $ here may be treated like a perturbation parameter. A brief
discussion behind this observation is given below, which leads to
more understanding the frame of Eq. (6).

Here, as a further knowledge to the reader, a general procedure is
outlined for the solution of Riccati equation in (6),
\begin{eqnarray}
\Delta W' _{n\ell } (r)=\frac{\sqrt{2m} }{\hbar } \left\{ \Delta W_{n\ell
}^{2} (r)+2W_{n\ell =0} (r)\Delta W_{n\ell }^{} (r)+\left[ \Delta \epsilon
_{n\ell } -\Delta V(r)\right] \right\}~,~~~~~(A.1)
\nonumber
\end{eqnarray}
from which the exact solution for $\Delta W$ reads
\begin{eqnarray}
\Delta W_{n\ell } (r)=\Delta W_{n\ell }^{sp} (r)-\frac{\exp \left[
2\frac{\sqrt{2m} }{\hbar } \int\limits_{}^{r}\left( \Delta W_{n\ell }^{sp}
(z)+W_{n\ell =0} (z)\right) dz \right] }{\frac{\sqrt{2m} }{\hbar }
\int\limits_{}^{r}\exp \left[ 2\frac{\sqrt{2m} }{\hbar }
\int\limits_{}^{y}\left( \Delta W_{n\ell }^{sp} (z)+W_{n\ell =0}
(z)\right) dz \right] dy }~,~~~~~(A.2)
\nonumber
\end{eqnarray}
where $\Delta W_{n\ell }^{sp} $ is a special solution of (A.1). More specifically, Eq. (A.2)
can be rewritten as
\begin{eqnarray}
\Delta W_{n\ell } (r)=\Delta W_{n\ell }^{sp} (r)-\frac{1/\left[ \phi
_{n\ell }^{sp} (r)\chi _{n\ell =0} (r)\right] ^{2}  }{\frac{\sqrt{2m}
}{\hbar } \int\limits_{}^{r}dz/\left[ \phi _{n\ell }^{sp} (z)\chi _{n\ell
=0} (z)\right] ^{2}   }~,~~~~~~~~~~~~(A.3)
\nonumber
\end{eqnarray}
or
\begin{eqnarray}
\Delta W_{n\ell } (r)=\Delta W_{n\ell }^{sp} (r)-\frac{\hbar }{\sqrt{2m}
} \frac{d}{dr} \ln \left\{ \int\limits_{}^{r}dz/\left[ \phi _{n\ell }^{sp}
(z)\chi _{n\ell =0} (z)\right] ^{2}   \right\}~,~~~~~~(A.4)
\nonumber
\end{eqnarray}
which is obvious that due to the present consideration of exactly solvable
potentials, the full superpotential in the above equations should involve
the whole of correction terms in a perturbation series discussed in detail below.

As $\chi _{n\ell =0} $ is known explicitly, we briefly discuss here the physics behind
$\Delta W_{n\ell }^{sp} $ leading to $\phi _{n\ell }^{sp} $
to find an exact analytical solution for $\Delta W_{n\ell } $. Using
the spirit of recently introcuded supersymmetric perturbation theory [5],
we expand the related functions in terms of $\ell $
that is treated here like a perturbation parameter,
\begin{eqnarray}
\Delta V(r;\ell )=\sum\limits_{k=1}^{\infty }\ell ^{k} \Delta V^{k}
(r)\;,\;\Delta W_{n\ell } (r)= \sum\limits_{k=1}^{\infty }\ell ^{k}
\Delta W_{n}^{k} (r)\;,\;\Delta \epsilon _{n\ell }
=\sum\limits_{k=1}^{\infty }\ell ^{k} \epsilon _{n}^{k}~.~~~~~~(A.5)
\nonumber
\end{eqnarray}

Substitution of the above expansion into (A.1) by equating terms
with the same power of $\ell $ on both sides yields up to
$O\left( \ell ^{2} \right) $, one arrives at
\begin{eqnarray}
2W_{n\ell =0} \Delta W_{n\ell }^{k=1} -\frac{\hbar }{\sqrt{2m} } \left(
\Delta W_{n\ell }^{k=1} \right) ^{} =\Delta V^{k=1} -\Delta \epsilon
_{n\ell }^{k=1}~,
\nonumber
\end{eqnarray}
\begin{eqnarray}
\left( \Delta W_{n\ell }^{k=1} \right) ^{2} +2W_{n\ell =0} \Delta
W_{n\ell }^{k=2} -\frac{\hbar }{\sqrt{2m} } \left( \Delta W_{n\ell }^{k=2}
\right) ^{} =\Delta V^{k=2} -\Delta \epsilon _{n\ell }^{k=2}~.~~~~~~(A.6)
\nonumber
\end{eqnarray}

From this short discussion, one sees that $\Delta W_{n\ell }^{k=1}
$ or $\Delta W_{n\ell }^{k=2} $ corresponding to the first and
second order, are the candidates for a special solution ($\Delta
W_{n\ell }^{sp} $) of (A.1) or (A.3). For simplicity, we choose
$\Delta W_{n\ell }^{k=1} $ and, from (A.6), give its explicit form
as
\begin{eqnarray}
\Delta W_{n\ell }^{k=1} (r)=\frac{\sqrt{2m} }{\hbar } \frac{1}{\left[
\chi _{n\ell =0} (r)\right] ^{2} } \int\limits_{-\infty }^{r}\left[ \chi
_{n\ell =0} (z)\right] ^{2} \left( \epsilon _{n\ell }^{k=1} -V^{k=1}
(z)\right)  dz~,~~~~~~(A.7)
\nonumber
\end{eqnarray}
in which $\Delta \epsilon _{n\ell }^{k=1} $ is evaluated by
\begin{eqnarray}
\Delta \epsilon _{n\ell }^{k=1} =N_{n\ell =0}^{2} \int\limits_{-\infty
}^{\infty}\left[ \chi _{n\ell =0} (z)\right] ^{2}  V^{k=1} (z)dz~.~~~~~~(A.8)
\nonumber
\end{eqnarray}

For instance, if one performs the calculations for the first
example discussed in this letter, that is the harmonic oscillator
potential with the angular momentum barrier worked out for $n=0$, then
it is not difficult to see that $\Delta \epsilon _{0\ell }^{k=1} =\ell \hbar w$
and higher order corrections are zero due to
\begin{eqnarray}
\Delta W_{0\ell }^{k=2} (r) & = & \frac{\sqrt{2m} }{\hbar } \frac{1}{\left[
\chi _{00} (r)\right] ^{2} } \int\limits_{-\infty }^{r}\left[ \chi _{00}
(z)\right] ^{2} \left[ \epsilon _{0\ell }^{k=2} +\left( \Delta W_{0\ell}^{k=1}(z)
\right)^{2}-V^{k=2}(z)\right]dz=0~,
\nonumber
\\
\Delta \epsilon _{0\ell }^{k=2} & = & N_{00}^{2} \int\limits_{-\infty
}^{\infty}\left[ \chi _{00} (z)\right] ^{2} \left[ V^{k=2} (z)-\left( \Delta
W_{0\ell }^{k=1} \left( z\right) \right) ^{2} \right]  dz=0~,~~~~~~(A.9)
\nonumber
\end{eqnarray}
because $V^{k=1} (r)=\ell \hbar ^{2} /2mr^{2}  $ and
$V^{k=2} (r)=\left( \ell \hbar \right) ^{2} /2mr^{2}  $, which leads to
$\Delta W_{0\ell }^{sp} (r)=\Delta W_{0\ell }^{k=1} (r)=-\ell \hbar
/\sqrt{2m} r $ that satisfies the first order expansion in (A.6).

\medskip \medskip
The author wish to thank the referee for his helpful comments and
suggestions.

\end{document}